\documentclass[12pt,reqno]{article}
\usepackage{amsfonts}
\usepackage{amsmath}
\usepackage{amsbsy}
\usepackage{graphicx}
\usepackage{amssymb,latexsym}
\numberwithin{equation}{section}
\usepackage{cite}

\begin{document}

\title{A note on the pp-wave solution of Minimal Massive 3D Gravity coupled with Maxwell-Chern-Simons theory }
\begin{titlepage}
\author{ Hakan Cebeci \footnote{E.mail: hcebeci@eskisehir.edu.tr  , 
hcebeci@gmail.com} \\ 
{\small Department of Physics, Eski\c{s}ehir Technical University, 26470 Eski\c{s}ehir, Turkey}  \\  \\
Tekin Dereli \footnote{E.mail: tdereli@ku.edu.tr , tekindereli@maltepe.edu.tr} \\
{\small Department of Basic Sciences, Faculty of Engineering and Natural Sciences,} \\ 
{\small Maltepe University, 34857 Maltepe, \.{I}stanbul, Turkey } \\
{\small Emeritus Professor of Physics, Ko\c{c} University, 34450 Sar{\i}yer, \.{I}stanbul, Turkey} \\  \\
Se\c{c}il \c{S}entorun \footnote {E.mail: secilo@eskisehir.edu.tr} \\
{\small Department of Physics, Eski\c{s}ehir Technical University, 26470 Eski\c{s}ehir, Turkey} }

\date{ }

\maketitle

\bigskip

\begin{abstract}
\noindent In this work, we examine a family of $pp$-wave solutions of Minimal Massive 3d Gravity (MMG) minimally coupled with the Maxwell-Chern-Simons theory. An elaborate investigation of the field equations shows that the theory admits $pp$-wave solutions provided that there exist an anti-self duality relation between the electric and the magnetic components of the Maxwell 2-form field. By employing Noether-Wald formalism, we also construct Noether charges of the theory within exterior algebra formalism.    

\vspace{0.4cm}

\end{abstract}

\end{titlepage}

\section{Introduction}

3-dimensional massive gravity models have attracted too much attention in recent years since the Einstein's theory of gravity in $ (2+1) $ dimensions has no propagating degrees of freedom. Adding a gravitational Chern-Simons term to the Einstein-Hilbert action with the wrong sign leads to a propagating theory, namely Topologically Massive Gravity (TMG), having third order gravitational field equations \cite{deser_1,deser_2}. Indeed, it has been shown that TMG contains one massive propagating graviton mode when a linearization of the field equations about $ AdS_{3} $ vacuum is made. However, the holographical methods illustrated in \cite{brown} have proven that although the bulk graviton mode in TMG has positive energy, one of the central charges calculated through Virasoro algebra turns out to be negative leading to a non-unitary holographically dual conformal field theory (CFT) on the 2d boundary \cite{kraus}. To overcome this bulk versus boundary clash problem, a new massive gravity theory, identified as New Massive Gravity or shortly NMG, has been proposed in \cite{bergshoeff_1} by introducing parity-even quadratic curvature terms in the action and discarding parity-odd topological Chern-Simons term of TMG. Yet, it has been seen that NMG has not satisfactorily resolved bulk-boundary clash problem of TMG. A generalization of NMG, the so-called Generalized Massive Gravity, has been introduced in \cite{bergshoeff_2} by adding parity odd gravitational Chern-Simons term to NMG action with the hope to fix the bulk versus boundary CFT unitarity problem. Recently, as a modification of TMG, Minimal Massive Gravity (MMG) has been introduced in \cite{bergshoeff_3} showing that MMG remarkably solves bulk-boundary clash problem within a certain range of its parameters \cite{arvanitakis_1}. It has been seen that MMG has same gravitational degrees of freedom as TMG and it also involves one massive bulk graviton. It has been further illustrated that bulk graviton has positive energy while there exist positive central charges for dual CFT leading to a unitary theory. 

Since similar physical features have been noted in TMG and MMG, both theories possess the similar type of vacuum solutions. It has been recently discovered that the most of the vacuum solutions of TMG (see \cite{chow_1,chow_2}) are also vacuum solutions of MMG. Interesting vacuum solutions of MMG include Kaluza-Klein $ AdS_{2} \times S^{1} $ type black hole solutions in addition to warped $ AdS_{3} $ vacuum solutions presented in \cite{arvanitakis_2}. There exist wave and warped AdS solutions discussed in \cite{alishahiha} where it is also proven that MMG admits logarithmic type solutions at the critical points. In \cite{arvanitakis_3}, conformally-flat vacuum solutions of MMG have been introduced where the Cotton tensor vanishes for such spacetimes. In addition, regarding Segre-Petrov classification, it has been illustrated that MMG also admits Petrov-type $ O $, $ D $ and type $ N $ spacetime solutions as TMG \cite{altas}. In addition, the existence of Kundt solutions have been investigated in \cite{deger_1} while in a recent work \cite{deger_2}, some different classes of homogeneous solutions including stationary Lifshitz type solution with Lifshitz parameter $ z = - 1 $ have been discovered. Finally in \cite{sarioglu}, static and stationary circularly symmetric solutions of MMG have been constructed at the merger point.

On the other hand, the extension of MMG with matter couplings deserves a particular investigation. Minimal matter coupling in MMG has been first accomplished in \cite{arvanitakis_2}. In that work, the authors have explicitly constructed the source tensor associated with MMG gravitational field equation considering that the matter Lagrangian is independent of the connection fields.  As a remarkable feature of matter-coupled MMG, it has been illustrated that the source tensor is quadratic in stress-energy tensor and it also includes the covariant derivative term related with the stress-energy tensor. In this respect, the matter coupling in MMG completely differs from matter coupling in TMG where matter-coupled gravitational field equation of TMG includes only the matter stress-energy tensor as a source. Within this framework, some exact solutions of matter-coupled MMG have also been investigated for certain matter Lagrangians. Referring to matter-coupled solutions, first in \cite{arvanitakis_2}, Friedmann-Lemaitre-Robertson-Walker (FLRW) type cosmological solutions have been constructed for ideal fluid. Later in \cite{arvanitakis_3}, the black string and domain wall solutions have been investigated for scalar matter Lagrangian involving minimally coupled dynamical scalar field with a potential term. Finally in a recent work \cite{nam_1}, warped $ AdS_{3}$ type black hole solutions have been discussed for MMG theory minimally coupled with Maxwell-Chern-Simons Lagrangian where in that work an interesting polynomial ansatz has been employed for the electromagnetic potential.  

In this work, motivated by \cite{dereli_3}, we aim to investigate the $pp$-wave type solutions of MMG minimally coupled with Maxwell-Chern-Simons theory within the exterior algebra formalism. It should also be noted that using the same formalism, MMG has been studied in the works \cite{baykal,dereli_1,dereli_2,dereli_4,cebeci} where a detailed investigation of the matter-coupling in the exterior algebra formalism is performed in the work \cite{cebeci}.  
Now in the first part of the present study, we obtain the field equations by making variations of the matter-coupled MMG Lagrangian with respect to co-frames, connections, auxiliary 1-form and Maxwell potential 1-form fields. By solving auxiliary field analytically in terms of Maxwell stress-energy 2-forms, we obtain the gravitational field equation of MMG with the source term that is quadratic in stress-energy forms. In the second part, we try to obtain $pp$-wave solutions of the gravitational field equation. We explicitly show that MMG coupled with Maxwell-Chern-Simons admits $pp$-wave solutions provided that the electric and the magnetic components of the Maxwell field satisfy an anti-self-duality condition. We examine the solutions for certain special cases. In the next part, by employing Noether-Wald formalism, we construct Noether charges within exterior algebra formalism. Finally, we end up our discussion with some closing remarks.

\section{MMG minimally coupled with Maxwell-Chern-Simons theory}
\label{section_1}

\noindent In this section, following \cite{cebeci}, let us first introduce the MMG theory coupled with Maxwell-Chern-Simons theory and the formalism used. In the language of differential forms, the model can be described by the action
\begin{equation}
I = \int_{M} \, L = \int_{M}\, ( L_{MMG} +  L_{E} )
\label{eqn_1_1}
\end{equation}
where the Lagrangian 3-form of Minimal Massive Gravity can be expressed in the form  
\begin{eqnarray}
& & L_{MMG} = - \frac{\sigma}{2}  R^{ab} \wedge \ast ( e_{a} \wedge e_{b} ) + \Lambda \ast 1    \nonumber \\ 
& & + \frac{1}{2 \mu} \left( \omega^{a}\,_{b} \wedge d \omega^{b}\,_{a} + \frac{2}{3} \omega^{a}\,_{b} \wedge \omega^{b}\,_{c} \wedge \omega^{c}\,_{a} \right) + \lambda_{a} \wedge T^{a} \nonumber \\ 
& & + \frac{\alpha}{2} \lambda_{a} \wedge \lambda_{b} \wedge \ast ( e^{a} \wedge e^{b} ) 
\label{eqn_1_2}
\end{eqnarray}
and the Maxwell-Chern-Simons Lagrangian 3-form reads
\begin{equation}
L_{E} = - \frac{ 1 }{2} F \wedge \ast F  - \frac{ 1 }{2} m A \wedge F  
\label{maxwell}
\end{equation}
where the Maxwell field $ F = d A $. We note that the basic field variables in the total Lagrangian are the co-frame 1-forms $ e^{a} $, in terms of which the 3-dimensional spacetime metric can be decomposed as $ g =\eta_{ab}\, e^{a} \otimes e^{b} $ with $\eta_{ab} = diag ( - + + )\,$, the connection 1-forms $ \omega^{a}\,_{b} $, the auxiliary 1-form $\lambda_{a}$ and the Maxwell potential 1-form $ A $. In MMG Lagrangian, $ \Lambda $ describes the cosmological constant. $ \mu $ denotes the mass parameter of TMG while $ m $ is the mass parameter associated with the Maxwell field. $ \alpha $ and $ \sigma $ are arbitrary dimensionless parameters that may take either sign.\footnote{We notice that in \cite{bergshoeff_3}, $ \sigma $ was taken as to satisfy $ \sigma^{2} = 1 $. In the present work, we will keep it arbitrary as in \cite{arvanitakis_2} and \cite{cebeci}.} Recall that the case $ \alpha = 0 $ results in the Lagrangian of TMG coupled with Maxwell-Chern-Simons theory where in that case the auxiliary 1-form field $ \lambda_{a} $ can be identified as the Lagrange multiplier 1-form that imposes the zero-torsion constraint.  
In addition, we note that $\ast$ is the Hodge operator which maps a $ p$-form to a $ (3-p)$-form in 3 dimensions. On the other hand, the oriented $3$-dimensional volume element can be written in the form $ \ast 1 = e^{0} \wedge e^{1} \wedge e^{2} $.
Furthermore, the connection 1-forms $ \omega^{a}\,_{b} $ obey the first Cartan structure equations
\begin{equation}
d e^{a} + \omega^{a}\,_{b} \wedge  e^{b} = T^{a}
\label{eqn_1_3}
\end{equation}
where $T^{a}$ identifies the torsion 2-forms while the
corresponding curvature 2-forms can be calculated from the second Cartan
structure equations
\begin{equation}
R^{a b} = d \omega^{a b} + \omega^{a}\,_{c} \wedge \omega^{c\,b}\,\, . 
\label{eqn_1_4}
\end{equation}
We also consider that the spacetime is metric-compatible which implies that $\omega_{a \,b} = - \omega_{b \,a}\, $.
For future use, we also note that $R=\iota_{a} P^{a}$ is the scalar curvature that can be expressed in terms of Ricci 1-forms $P^{a} = \iota_{b} R^{ba} $ where $\iota_{b}$ is the inner product operator that respects the identity $\iota_{b} e^{a} = \delta_{b}^{a} $. In addition, we remark that the dreibeins $ e^{a}\,_{\mu} $ obtained through $ e^{a} = e^{a}\,_{\mu} d x^{ \mu} $ are assumed to be invertible. To obtain the field equations, we consider independent variations of the total action with respect to co-frames $e^{a}$, connection 1-forms $\omega^{a}\,_{b}$, the auxiliary 1-form $ \lambda_{a} $ and Maxwell $1$-form potential $ A $.   
Then an arbitrary variation of the action results in
\begin{equation}
\delta I = \int_{M} \, \delta L = \int_{M} ( \delta L_{MMG} + \delta  L_{E} ) 
\end{equation}
where the variation of total Lagrangian reads
\begin{eqnarray}
\delta L &=& \delta e^{c} \wedge \left( - \frac{ \sigma }{2} R^{ab} \wedge \ast ( e_{a} \wedge e_{b} \wedge e_{c} ) + \Lambda \ast e_{c} + D \lambda_{c} + \frac{ \alpha }{2} \epsilon_{abc} \lambda^{a} \wedge \lambda^{b} + t_{c} \right) \nonumber \\
& & + \delta \omega_{ab} \wedge \left( - \frac{ \sigma }{2} D ( \ast ( e^{a} \wedge e^{b} ) ) + \frac{ 1 }{\mu} R^{ba} - \frac{ 1 }{2} ( \lambda^{a} \wedge e^{b} - \lambda^{b} \wedge e^{a} ) \right) \nonumber \\
& & + \delta \lambda_{a} \wedge \left( T^{a} + \alpha \lambda_{b} \wedge \ast ( e^{a} \wedge e^{b} ) \right)  - \delta A \wedge ( d ( \ast F ) + m F ) + d \theta \, .
\label{variation}
\end{eqnarray}
$ \theta $ denotes the surface term obtained from variation in the form
\begin{eqnarray}
\theta = - \frac{ \sigma }{2} \delta \omega_{ab} \wedge \ast ( e^{a} \wedge e^{b} ) + \frac{ 1 }{2 \mu} \delta \omega_{ba} \wedge \omega^{ab} \nonumber \\
+ \delta e^{a} \wedge \lambda_{a} - \delta A \wedge \ast F - \frac{ 1 }{2} m \delta A \wedge A \, . 
\label{boundary_term}
\end{eqnarray}
Then from the variation, one gets the following field equations :
\begin{equation}
E_{c} : = - \frac{\sigma}{2} \epsilon_{abc} R^{ab} + \Lambda \ast e_{c} + D \lambda_{c} + \frac{\alpha}{2} \epsilon_{abc} \lambda^{a} \wedge \lambda^{b}  + t_{c} = 0 \, ,
\label{eqn_1_5}
\end{equation}
\begin{equation}
E^{ab} : = - \frac{\sigma}{2} D ( \ast ( e^{a} \wedge e^{b} ) ) + \frac{1}{ \mu } R^{ba} - \frac{1}{2} ( \lambda^{a} \wedge e^{b} - \lambda^{b} \wedge e^{a} ) = 0 \, ,
\label{eqn_1_6}
\end{equation}
\begin{equation}
N^{a} : = T^{a} + \alpha \lambda_{b} \wedge \ast ( e^{a} \wedge e^{b} ) = 0 
\label{eqn_1_7}
\end{equation}
together with Maxwell field equation
\begin{equation}
N : = d ( \ast F ) + m F = 0
\label{maxwell_equation} \, . 
\end{equation}
Note that in obtaining field equations, we have employed 3-dimensional identities
\begin{equation}
\ast ( e_{a} \wedge e_{b} ) = \epsilon_{abc} e^{c} \, , \qquad \ast ( e_{a} \wedge e_{b} \wedge e_{c} ) = \epsilon_{abc}  \, .
\label{eqn_1_8}  
\end{equation}
In addition, we notice that $ t_{c} $ denotes the stress-energy 2-form associated with Maxwell 2-form field $ F $ expressed as
\begin{equation}
t_{c} = \frac{ \delta L_{E} }{ \delta e^{c}}  = \frac{ 1 }{2} ( \iota_{c} F \wedge \ast F - F \wedge \iota_{c} ( \ast F ) ) \, .
\label{stress_energy}
\end{equation}
Then using (\ref{eqn_1_7}) and the relation
\begin{equation}
K^{a}\,_{b} \wedge e^{b} = T^{a} \, ,
\label{eqn_1_9}
\end{equation}
one can determine the contorsion one forms $K^{a}\,_{b} $ in the form
\begin{equation}
K^{a}\,_{b} = \alpha \epsilon^{a}\,_{bc} \lambda^{c} \, . 
\label{contorsion}
\end{equation}
Recall that the equations (\ref{eqn_1_5}) and (\ref{eqn_1_6}) include connection $ \omega^{a}\,_{b}$ terms with torsion. In order to get an algebraic solution for the auxiliary 1-form $ \lambda_{a} $, it requires that these equations should be expressed in terms of torsion-free connection. For that reason, it is customary to decompose the connection one-form $ \omega^{a}\,_{b} $ as
\begin{equation}
\omega^{a}\,_{b} = \bar{\omega}^{a}\,_{b} + K^{a}\,_{b} 
\label{connection_decomposition}
\end{equation}
in terms of torsion-free (Levi-Civita) connections $ \bar{\omega}^{a}\,_{b} $ and contorsion one-forms $ K^{a}\,_{b} $. Then if one uses Cartan structure equation (\ref{eqn_1_4}), one can similarly express curvature two forms $R^{ab} $ 
\begin{equation}
R^{ab} = \bar{R}^{ab} + \bar{D} K^{ab} + K^{a}\,_{c} \wedge K^{cb}
\label{curvature_decomposition}
\end{equation}
in terms of curvature two forms $ \bar{R}^{ab} $ of torsion-free connection $ \bar{\omega}^{a}\,_{b} $ and contorsion and $ \bar{D} $ denotes the exterior covariant derivative operator acting with respect to torsion-free connection. Then using the expression (\ref{contorsion}), the curvature 2-form reads 
\begin{equation}
R^{ab} = \bar{R}^{ab} +  \alpha \epsilon^{abc} \bar{D} \lambda_{c} + \alpha^{2} \lambda^{a} \wedge \lambda^{b}  \, . 
\label{eqn_1_10}
\end{equation} 
in terms of the curvature 2-form associated with the torsion-free connection and the auxiliary 1-form $ \lambda^{a} $. \footnote{ We refer reader to \cite{cebeci} for detailed calculations. } 
Finally, substituting (\ref{eqn_1_10}) into equations (\ref{eqn_1_5}) and (\ref{eqn_1_6}), one derives the following field equations expressed in terms of torsion-free connection 
\begin{equation}
- \frac{\sigma }{2} \epsilon_{abc} \bar{R}^{ab} + ( \alpha \sigma + 1 ) \bar{D} \lambda_{c} - \frac{1}{2} \alpha ( \alpha \sigma + 1 ) \epsilon_{abc} \lambda^{a} \wedge \lambda^{b} + \Lambda \ast e_{c} + t_{c} = 0 
\label{einstein_equation}  
\end{equation}
and
\begin{equation}
- \frac{1}{2} ( \alpha \sigma + 1 ) ( \lambda^{a} \wedge e^{b} - \lambda^{b} \wedge e^{a} ) + \frac{1}{\mu} ( \bar{R}^{ba} + \alpha \, \epsilon^{bac} \bar{D} \lambda_{c} + \alpha^{2} \lambda^{b} \wedge \lambda^{a} ) = 0 
\label{connection_equation}
\end{equation}
where it is also imposed that $ \alpha \sigma + 1 \neq 0 $.  
To get an algebraic solution for the auxiliary field, the terms $ \alpha \epsilon^{bac} \bar{D} \lambda_{c} $ and $ \alpha^{2} \lambda^{b} \wedge \lambda^{a} $ present in the connection field equation (\ref{connection_equation}) should be eliminated. This can be accomplished by using Einstein field equation (\ref{einstein_equation}). As a result, the equation for $ \lambda_{a} $ simply becomes
\begin{equation}
\lambda^{a} \wedge e^{b} - \lambda^{b} \wedge e^{a} = \frac{2}{\mu ( \alpha \sigma + 1 )^{2} } ( \bar{R}^{ba} - \alpha \epsilon^{bac} t_{c} + \alpha \Lambda e^{b} \wedge e^{a} ) 
\label{eqn_1_14}
\end{equation}  
whose solution reads (referring to \cite{cebeci})
\begin{equation}
\lambda^{a} = - \frac{2}{ \mu ( \alpha \sigma + 1 )^{2} } \left( \bar{Y}^{a} + \frac{ 1 }{2} \alpha \Lambda e^{a} + \alpha \epsilon^{abc} \iota_{b} t_{c} - \frac{1}{4} \alpha \epsilon^{npc} \iota_{n} ( \iota_{p} t_{c} ) e^{a} \right) 
\label{lambda_1} 
\end{equation}
where
\begin{equation}
\bar{Y}^{a} = \bar{P}^{a} - \frac{ 1 }{4} \bar{R} e^{a}
\label{schouten}
\end{equation}
denotes Schouten 1-forms expressed in terms of torsion-free Ricci 1-forms $ \bar{P}^{a} $ and the curvature scalar $ \bar{R} $. 
In terms of the Maxwell field $ F $, $ \lambda^{a} $ can be expressed in the form
\begin{equation}
\lambda^{a} = - \frac{ 2 }{ \mu ( \alpha \sigma + 1 )^{2} } \left( \bar{Y}^{a} + \frac{ 1 }{2} \Lambda e^{a} + \frac{ 1 }{2} \alpha \epsilon^{abc} F_{cb} \ast F + \frac{ 1 }{4} \alpha \ast( F \wedge \ast F ) e^{a} \right) \, .
\label{lambda_2}
\end{equation}  
For later purposes, it would also be beneficial to express auxiliary 1-form $ \lambda^{a} $ alternatively in the simplified form
\begin{equation}
\lambda^{a} = - \frac{2}{ \mu ( \alpha \sigma + 1 )^{2} } \left( \bar{Y}^{a} + \frac{ 1 }{2} \alpha \Lambda e^{a} + \alpha \ast \hat{t}^{a} \right) \, . 
\label{lambda_3}
\end{equation} 
where the new stress-energy 2-forms is defined as
\begin{equation}
\hat{t}_{c} = t_{c} - \frac{ 1 }{2} t \ast e_{c} 
\end{equation}
in terms of the trace $ t $ of the Maxwell stress-energy 2-form. Also note that the trace reads
\begin{equation}
t = - \frac{ 1 }{ 2 } \ast ( F \wedge \ast F )  \, .
\label{trace}
\end{equation}
Then, if one substitutes (\ref{lambda_3}) into Einstein field equation (\ref{einstein_equation}) expressed in terms of torsion-free connection, one obtains the field equation of MMG theory minimally coupled with Maxwell-Chern-Simons. After simplifications, one finally gets 
\begin{equation}
\alpha_{1} \epsilon_{abc} \bar{R}^{ab} + \alpha_{2} \bar{C}_{c} + \alpha_{3} \ast e_{c} + \frac{ 1 }{2} \alpha_{4} \epsilon_{abc} \bar{Y}^{a} \wedge \bar{Y}^{b} + \tilde{t}_{c}= 0
\label{mmg_equation}
\end{equation}
where 
\begin{equation}
\bar{C}_{c} = \bar{D} \bar{Y}_{c}
\label{cotton}
\end{equation}
denotes Cotton 2-forms with respect to torsion-free connection and the coefficients read
\begin{eqnarray}
\alpha_{1} = - \frac{ 1 }{2} \left( \sigma + \frac{ 2 \alpha^{2} \Lambda }{ \mu^{2} ( \alpha \sigma + 1 )^{3}} \right) \qquad \alpha_{2} = - \frac{2}{ \mu ( \alpha \sigma + 1 ) } \nonumber \\
\alpha_{3} = \Lambda - \frac{ \alpha^{3} \Lambda^{2} }{ \mu^{2} ( \alpha \sigma + 1 )^{3} } \qquad \alpha_{4} = - \frac{ 4 \alpha }{ \mu^{2} ( \alpha \sigma + 1 )^{3} } \, .
\label{alpha}
\end{eqnarray}
Also remark that while obtaining (\ref{mmg_equation}) we use the identity 
\begin{equation}
\bar{R}^{ab} = \bar{Y}^{a} \wedge e^{b} - \bar{Y}^{b} \wedge e^{a} \, .
\label{curvature_schouten_identity}
\end{equation}
On the other hand, the source term $ \tilde{t}_{c} $ explicitly reads
\begin{eqnarray}
\tilde{t}_{c} &=& \left( 1 - \frac{ 2 \alpha^{3} \Lambda }{ \mu^{2} ( \alpha \sigma + 1 )^{3} }  \right) t_{c} - \frac{ 2 \alpha^{2} }{\mu^{2} ( \alpha \sigma + 1 )^{3} } \epsilon_{abc} ( \bar{Y}^{a} \wedge \ast \hat{t}^{b} - \bar{Y}^{b} \wedge \ast \hat{t}^{a} ) \nonumber \\
& & - \frac{ 2 \alpha }{ \mu ( \alpha \sigma + 1 ) } \bar{D} ( \ast \hat{t}_{c} ) - \frac{ 2 \alpha^{3} }{ \mu^{2} ( \alpha \sigma + 1 )^{3} } \epsilon_{abc} \ast \hat{t}^{a} \wedge \ast \hat{t}^{b} 
\label{source_term}
\end{eqnarray} 
expressed in terms of redefined Maxwell stress-energy 2-forms $ \hat{t}_{c} $. We further note that the source includes terms that are quadratic in stress-energy forms as is also discussed in \cite{arvanitakis_2} and \cite{cebeci}. It also involves the exterior covariant derivative of redefined Maxwell stress-energy 2-form. Another important issue that deserves further discussion is the consistency of MMG field equation. Indeed, it has been shown in the previous works that the consistency of MMG field equation requires the source term to satisfy a relation of the form (expressed in terms of differential forms)  
\begin{equation}
\bar{D} \tilde{ t }_{c} = \frac{ \alpha_{4} }{ \alpha_{2} } \epsilon_{abc} \tilde{ t }^{a} \wedge \bar{Y}^{b} \, .
\label{consistency}
\end{equation}
\footnote{Note that this expression is analogous to the expression (4.15) in \cite{arvanitakis_2} presented in the tensor notation.} Note that this consistency relation holds for any matter source obtained from the connection-independent matter coupling. The equality of left and right hand sides of the consistency expression for any connection-independent matter coupling has been explicitly illustrated in \cite{cebeci} by using the exterior algebra formalism. This has been performed explicitly by recalling that stress-energy 2-forms $ t_{c} $ are covariantly conserved i.e $ \bar{D} t_{c} = 0 $ and noting that identities
\begin{equation} 
\bar{D}^{2} ( \ast \hat{ t }_{c} ) = \bar{R}_{cb} \wedge \ast \hat{ t }^{b}  \, ,
\end{equation} 
\begin{equation}
\bar{Y}_{c} \wedge e_{b} \wedge \ast \hat{ t }^{b} = 0 
\end{equation}
and
\begin{equation}
\epsilon_{abc} t^{a} \wedge \bar{Y}^{b} = \bar{Y}^{b} \wedge e_{c} \wedge \ast t_{b} 
\end{equation}
hold. It can be seen that the left and the right hand sides of the consistency relation become equal for arbitrary values of $ \Lambda $.

\section{$pp$-wave solutions }
\label{section_2}

\noindent In this part, we investigate $pp$-wave solutions of the MMG field equation (\ref{mmg_equation}). To this end, we choose the local coordinates $(u,v,\rho)$ and take the $pp$-wave metric in the form
\begin{equation}
g = f (u, \rho) d u^{2} + 2 e^{ 2 c \rho } du \, dv + d \rho^{2}
\label{metric} 
\end{equation}
where the metric function $ f(u, \rho )$ and the exponential constant $c$ should be determined from the field equations. In general, the Maxwell 2-form field should involve two electric components and one magnetic component in 3-dimensions. However, in \cite{blagojevic_1} it has been shown that for the static and spherically symmetric solutions of 3-dimensional Einstein-Maxwell theory with gravitational Chern-Simons term, one of the components of Maxwell field has to vanish. Furthermore, if one looks for the self-dual solutions of 3-dimensional gravity theories (with and without torsion) coupled with Maxwell-Chern-Simons theory, the Maxwell field involves one electric and one magnetic component in general. Our aim is to obtain self-dual solutions of MMG coupled with Maxwell-Chern-Simons theory. Therefore motivated by the works \cite{dereli_3}, \cite{dereli_5} and \cite{blagojevic_2}, we consider the ansatz  
\begin{equation}
F = d A = E ( u , \rho ) e^{0} \wedge e^{1} + B ( u , \rho ) e^{1} \wedge e^{2} 
\label{maxwell}
\end{equation}
for Maxwell field assuming that the electric and magnetic parts depend on coordinate $ u $ and radial coordinate $ \rho $. Then, substituting the curvature, Schouten and Cotton forms for the $pp$-wave metric presented in the Appendix section together with the Maxwell field ansatz into (\ref{mmg_equation}), one obtains the following set of matter-coupled MMG gravitational field equations :
\begin{eqnarray}
- \alpha_{2} \beta^{\prime \prime} + \left( \frac{1}{2} \alpha_{4} c^{2} + \alpha_{2} c  - 2 \alpha_{1} \right) \beta^{\prime} = - \left( 1 - \frac{ 2 \alpha^{3} \Lambda }{\mu^{2} ( \alpha \sigma + 1 )^{3}} \right) EB \nonumber \\
+ \frac{ 2 \alpha }{ \mu ( \alpha \sigma + 1)} \left( (B^{2} + E B) ( \beta - c) - c E B - \frac{ 1}{2} ( E E^{\prime} + 3 B B^{\prime} )  \right) \nonumber \\
- \frac{ 2 \alpha^{2} }{ \mu^{2} ( \alpha \sigma + 1 )^{3} } \left( c^{2} E B + \frac{ 1 }{2} ( E^{2} - B^{2} ) \beta^{\prime} \right) + \frac{ \alpha^{3} }{\mu^{2} ( \alpha \sigma + 1)^{3} } ( E^{2} - B^{2} ) E B  \, ,
\label{mmg_1_a}
\end{eqnarray} 
\begin{equation}
\dot{E} \left( \frac{ 1 }{2} E + B \right) + \dot{B} \left( \frac{3}{2} B + E \right) = 0 \, ,
\label{mmg_1_b}
\end{equation}
\begin{eqnarray}
- \alpha_{2} \beta^{ \prime \prime} - 2 \alpha_{1} ( \beta^{\prime} + c^{2} ) + \alpha_{2} c \beta^{\prime} + \alpha_{3} + \frac{1}{2}  \alpha_{4} c^{2} \left( \beta^{\prime} + \frac{1}{2} c^{2} \right) = \nonumber \\
\frac{1}{2} \left( 1 - \frac{ 2 \alpha^{3} \Lambda }{ \mu^{2} ( \alpha \sigma + 1 )^{3} } \right) ( E^{2} + B^{2} ) + \frac{ 2 \alpha }{ \mu ( \alpha \sigma + 1 ) } \left( E^{\prime} B + E B^{\prime} \right. \nonumber \\
\left. + ( B^{2} + E B) ( \beta + c ) + c E^{2} \right) + \frac{ \alpha^{2} }{ \mu^{2} ( \alpha \sigma + 1 )^{3} } \left( \frac{ 1 }{2} c^{2} ( 3 E^{2} + B^{2} ) \right. \nonumber \\ 
\left. - ( E^{2} - B^{2} ) \left( \beta^{\prime} + \frac{ 1 }{2} c^{2} \right) \right) - \frac{ \alpha^{3} }{ 4 \mu^{2} ( \alpha \sigma + 1 )^{3}} ( E^{2} - B^{2} ) ( 3 E^{2} + B^{2} ) \, ,
\label{mmg_1_c}
\end{eqnarray}
\begin{equation}
E \dot{E} - B \dot{B} = 0 \, ,
\label{mmg_2_a}
\end{equation}
\begin{eqnarray}
\frac{ 1 }{4} \alpha_{4} c^{4} - 2 \alpha_{1} c^{2} + \alpha_{3} = \frac{ 1 }{2} \left( 1 - \frac{ 2 \alpha^{3} \Lambda }{ \mu^{2} ( \alpha \sigma + 1 )^{3} } \right) ( E^{2} - B^{2} ) \nonumber \\
- \frac{ 2 \alpha }{ \mu ( \alpha \sigma + 1 ) } ( E + B )^{2} \beta + \frac{ \alpha^{2} }{ \mu^{2} ( \alpha \sigma + 1 )^{3} } \left(  c^{2}  ( E^{2} - B^{2} ) - 4 ( E + B )^{2} \beta^{\prime} \right) \nonumber \\
- \frac{ 3 \alpha^{3} }{ 4 \mu^{2} ( \alpha \sigma + 1 )^{3} }  ( E^{2} - B^{2} )^{2} \, ,
\label{mmg_2_b}
\end{eqnarray}
\begin{eqnarray}
- \alpha_{2} \beta^{\prime \prime} + 2 \alpha_{1} ( c^{2} - \beta^{\prime} ) + \alpha_{2} c \beta^{\prime} - \alpha_{3} + \frac{ 1 }{2} \alpha_{4} c^{2} \left( \beta^{ \prime } - \frac{ 1 }{2} c^{2} \right) = \nonumber \\
\frac{ 1 }{2} \left( 1 - \frac{ 2 \alpha^{3} \Lambda }{ \mu^{2} ( \alpha \sigma + 1 )^{3} } \right) ( E^{2} + B^{2} ) + \frac{ 2 \alpha }{ \mu ( \alpha \sigma + 1 ) } \left( E^{\prime} B + E B^{\prime} \right. \nonumber \\
\left. - ( E^{2} + E B ) ( \beta - c ) + c B^{2} \right) + \frac{ \alpha^{2} }{ \mu^{2} ( \alpha \sigma + 1 )^{3} } \left( \frac{ 1 }{2} c^{2} ( E^{2} + 3 B^{2} ) \right. \nonumber \\
\left. - ( E^{2} - B^{2} ) \left( \beta^{ \prime } - \frac{ 1 }{2} c^{2} \right) \right) - \frac{ \alpha^{3} }{ 4 \mu^{2} ( \alpha \sigma + 1 )^{3} } ( E^{2} - B^{2} ) ( E^{2} + 3 B^{2} ) \, ,
\label{mmg_3_a}
\end{eqnarray}   
\begin{equation}
\dot{E} \left( \frac{3}{2} E + B \right) + \dot{B} \left( \frac{ 1 }{2} B + E \right) = 0 
\label{mmg_3_b}
\end{equation}
and
\begin{eqnarray}
- \alpha_{2} \beta^{ \prime \prime} + \left( \frac{ 1 }{2} \alpha_{4} c^{2} + \alpha_{2} c - 2 \alpha_{1} \right) \beta^{\prime} = - \left( 1 - \frac{ 2 \alpha^{3} \Lambda }{ \mu^{2} ( \alpha \sigma + 1 )^{3} } \right) E B \nonumber \\
- \frac{ 2 \alpha }{ \mu ( \alpha \sigma + 1 )} \left( ( E^{2} + E B ) ( \beta + c ) + c E B + \frac{ 1 }{2} ( 3 E E^{\prime} + B B^{ \prime } ) \right) \nonumber \\
- \frac{ 2 \alpha^{2} }{ \mu^{2} ( \alpha \sigma + 1 )^{3} } \left( c^{2} E B + \frac{ 1 }{2} ( E^{2} - B^{2} ) \beta^{\prime} \right) + \frac{ \alpha^{3} }{ \mu^{2} ( \alpha \sigma + 1 )^{3} } ( E^{2} - B^{2} ) E B    
\label{mmg_3_c}
\end{eqnarray}
where we have defined the function $ \beta=\beta(u,\rho)$ as 
\begin{equation}
\beta (u, \rho) = \frac{ 1 }{4} f^{\prime} - \frac{c}{2} f
\label{beta_definition}
\end{equation}
in terms of the metric function $ f$ and constant $ c$ and a prime denotes differentiation with respect to metric variable $ \rho $
while a dot denotes differentiation with respect to metric variable $ u$. Recall that the equations (\ref{mmg_1_b}), (\ref{mmg_2_a}) and (\ref{mmg_3_b}) that contain the partial derivatives of the Maxwell fields $ E$ and $ B $ with respect to $ u$ arise from the covariant derivative term in the source 2-form (\ref{source_term}). In fact, since we assume that the components of Maxwell 2-form field depend on the coordinates $ u $ and $ \rho $, covariant derivative in the source term in MMG field equation produces equations (\ref{mmg_1_b}), (\ref{mmg_2_a}) and (\ref{mmg_3_b}). Furthermore from (\ref{maxwell_equation}), Maxwell equations can be found as
\begin{equation}
E^{\prime} + ( \beta + c ) E + ( \beta + c - m ) B = 0 \, ,
\label{maxwell_1_a}
\end{equation}
\begin{equation}
\dot{E} + \dot{B} = 0 
\label{maxwell_1_b}
\end{equation}
and
\begin{equation}
B^{\prime} + ( c - \beta ) B + ( c - \beta - m ) E = 0 \, . 
\label{maxwell_1_c}
\end{equation}
From (\ref{maxwell_1_b}), one immediately obtains $ \dot{E} = - \dot{B} $ or $ \dot{B} = - \dot{E} $. Next substituting this in either of the equations (\ref{mmg_1_b}), (\ref{mmg_2_a}) and (\ref{mmg_3_b}) leads to $ E = - B $ or $ B = - E $ which implies that there should exist an anti-self-duality condition between the electric and magnetic components of the Maxwell 2-form field. Then inserting $ B = - E $ (and also $ B^{\prime} = - E^{\prime} $) in the remaining field equations, the system reduces to the following set of equations that are to be solved for the metric function $ f(u,\rho) $, the constant $ c$ and Maxwell field $ E$ : 
\begin{eqnarray}
- \alpha_{2} \beta^{ \prime \prime } + \left( \frac{ 1 }{2} \alpha_{4} c^{2} + \alpha_{2} c - 2 \alpha_{1} \right) \beta^{\prime} = \left( 1 - \frac{ 2 \alpha^{3} \Lambda }{\mu^{2} ( \alpha \sigma + 1 )^{3}} \right) E^{2} \nonumber \\
- \frac{ 2 \alpha }{ \mu ( \alpha \sigma + 1 ) } ( 2 E E^{\prime} - c E^{2} ) + \frac{ 2 \alpha^{2} }{ \mu^{2} ( \alpha \sigma + 1 )^{3}} c^{2} E^{2}  \, , 
\label{equation_metric}
\end{eqnarray}
\begin{equation}
\frac{ 1 }{4} \alpha_{4} c ^{4} - 2 \alpha_{1} c^{2} + \alpha_{3} = 0
\label{equation_constant}
\end{equation}
and
\begin{equation}
E^{ \prime} + m E = 0  \, .
\label{equation_maxwell}
\end{equation}
We also note the $B$-field equation 
\begin{equation}
B^{\prime} + m B = 0
\end{equation}
which is completely analogous to $E$-field equation. Note that (\ref{equation_maxwell}) can easily be solved in the form
\begin{equation}
E ( u, \rho ) = E_{0} (u) e^{ - m \rho }
\label{electric_field}
\end{equation}
where $ E_{0} (u) $ is an arbitrary function of its argument obtained from the integration with respect to metric variable $ \rho $. We also note that the $B$-field reads 
\begin{equation}
B(u , \rho ) = B_{0} ( u) e^{- m \rho}
\label{magnetic_field}
\end{equation}
where $ B_{0} ( u ) = - E_{0} (u) $ (i.e the anti-self-duality condition). Then, choosing the gauge 
\begin{equation}
A = A_{u} ( \rho , u ) \, d u
\label{maxwell_potential}
\end{equation}
for the Maxwell potential one-form and using $ F = d A $, the Maxwell one-form field can easily be determined in the form
\begin{equation}
A = \frac{ \sqrt{ 2 } }{ m } E_{0} e^{ - m \rho } d u 
\label{maxwell_potential_1}
\end{equation}
for $ m \neq 0 $ while for $ m = 0 $ it reads
\begin{equation}
A = - \sqrt{ 2 } E_{0} \rho \, d u
\label{maxwell_potential_2} 
\end{equation}
choosing the integration constant to be zero.

\noindent On the other hand, when using the expressions (\ref{alpha}) for $ \alpha_{i}$'s $ (i=1,2,3,4) $, the constant $ c $ can be determined from (\ref{equation_constant}) as
\begin{equation}
c^{2}_{\mp} = \frac{ \mu^{2} ( \alpha \sigma + 1 )^{3} }{ \alpha } \left( \frac{ 1 }{2} \left( \sigma + \frac{ 2 \alpha^{2} \Lambda }{ \mu^{2} ( \alpha \sigma + 1 )^{3} } \right) \pm \sqrt{ \frac{ 1 }{4} \sigma^{2} + \frac{ \Lambda \alpha }{ \mu^{2} ( \alpha \sigma + 1 )^{2}}} \right) 
\label{parameter_c}
\end{equation} 
in terms of the parameters of MMG. Furthermore considering the bulk and boundary unitarity conditions of MMG as discussed in \cite{arvanitakis_1}, it can be seen that the parameters of the theory are restricted to be in the ranges $ -1 < \alpha < 0 $ and $ \alpha \Lambda > 0 $. This choice makes inside the square root strictly positive. Then to be confident about positivity of $ c^{2} $, we choose the solution with plus sign (i.e the root $c_{+}^{2}$) as physical root. \footnote{Note that by taking $ \alpha = 0 $, $ \sigma = 1 $ and identifying $ \Lambda = - \frac{ 1 }{ \ell^{2} } $, one obtains $ c = \mp \frac{ 1 }{ \ell } $. In the absence of matter coupling, the solutions then correspond to vacuum AdS $pp$-wave solutions presented in \cite{chow_1}.}
Next substituting the electric field expression (\ref{electric_field}) into the MMG field equation (\ref{equation_metric}) and making some rearrangements, the MMG equation turns into 
\begin{equation}
\beta^{\prime \prime} - \mu_{1} \beta^{ \prime } = \nu_{1} E_{0}^{2} (u) e^{- 2 m \rho } 
\label{mmg_beta}
\end{equation} 
where we further identify the constants
\begin{eqnarray}
\mu_{1} &=& \frac{ 1 }{2} \frac{ \alpha_{4}}{\alpha_{2}} c^{2} + c - \frac{ 2 \alpha_{1}}{ \alpha_{2}} \nonumber \\ 
& = & \frac{ \alpha c^{2} }{ \mu ( \alpha \sigma + 1 )^{2} } + c - \frac{ \mu ( \alpha \sigma + 1 )}{2} \left( \sigma + \frac{ 2 \alpha^{2} \Lambda }{ \mu^{2} ( \alpha \sigma + 1 )^{3}} \right)
\label{mu_1}
\end{eqnarray}
and
\begin{eqnarray}
\nu_{1} = - \frac{ 1 }{ \alpha_{2} } \left( 1 - \frac{ 2 \alpha^{3} \Lambda }{ \mu^{2} ( \alpha \sigma +1 )^{3} } + \frac{2 \alpha }{ \mu ( \alpha \sigma + 1 ) } ( 2 m + c) + \frac{ 2 \alpha^{2} c^{2} }{ \mu^{2} ( \alpha \sigma + 1 )^{3}} \right) \nonumber \\ 
= \frac{ \mu ( \alpha \sigma + 1 ) }{2} \left( 1 + \frac{ 2 \alpha }{ \mu ( \alpha \sigma + 1 ) } (2m + c) + \frac{ 2 \alpha^{2} }{ \mu^{2} ( \alpha \sigma + 1 )^{3} } ( c^{2} - \alpha \Lambda ) \right) \, .
\label{nu_1}
\end{eqnarray}
Using (\ref{beta_definition}), MMG field equation can also be expressed in terms of the metric function $ f(u, \rho) $ in the form
\begin{equation}
f^{ \prime \prime \prime } - ( 2 c + \mu_{1} ) f^{ \prime \prime} + 2 \mu_{1} c f^{\prime} = 4 \nu_{1} E_{0}^{2} (u) e^{ - 2 m \rho } 
\label{mmg_metric_function}
\end{equation}
where the constants $ \mu_{1} $ and $ \nu_{1} $ are defined as in (\ref{mu_1}) and (\ref{nu_1}) respectively. Then for the non-homogeneous differential equation given above, we find the following $pp$-wave solutions for different special cases :  \\

\noindent {\bf 1.} The case $\mu_{1} \neq 2 c $, $ \mu_{1} \neq - 2 m $, $ c \neq - m $ and $ m \neq 0 $ : \\

\noindent For this general case, we obtain the metric solution
\begin{eqnarray}
f(u, \rho ) &=& - \frac{ \nu_{1} E_{0}^{2} (u) }{m (m+c) ( \mu_{1} + 2m) } e^{- 2m \rho } + \frac{ 4 C_{1} (u) }{ \mu_{1} ( \mu_{1} - 2 c)} e^{\mu_{1} \rho } \nonumber \\ 
& & - \frac{ 2 }{c} C_{2} (u) + C_{3} (u) e^{ 2 c \rho }
\label{pp_wave_1}
\end{eqnarray}
where $ C_{i} $'s ($i = 1,2,3)$) are arbitrary functions of their argument (that can also be identified as $pp$-wave profile functions) resulting from successive integrations with respect to metric variable $ \rho $. \\

\noindent {\bf 2.} The case $ \mu_{1} = 2 c $, $ \mu_{1} \neq - 2 m $, $ c \neq - m $ and $ m \neq 0 $ : \\

\noindent For this case, the solution can be given by
\begin{equation}
f( u , \rho ) = - \frac{ \nu_{1} E_{0}^{2} (u)}{ 2 m ( m +c)^{2} } e^{ - 2 m \rho } + \frac{ 2 }{c} C_{1} (u) \rho \, e^{ 2 c \rho } - \frac{2}{c} C_{2} (u) + C_{3} (u) e^{2 c \rho } \, . 
\label{pp_wave_2}
\end{equation}  
Recall that the solution is valid for the condition $ \mu_{1} = 2 c $ which implies the relation
\begin{equation}
\frac{ \sigma^{2} }{ 16 } ( \alpha \sigma + 2 )^{2} = \frac{ 1 }{4} \sigma^{2} ( \alpha \sigma + 1 )^{2} + \frac{ \Lambda \alpha }{\mu^{2}} 
\label{relation_1}
\end{equation}
between the parameters of MMG. \\

\noindent {\bf 3.} The case $ \mu_{1} \neq 2 c $, $ \mu_{1} = - 2 m $, $ c \neq - m $ and $ m \neq 0 $ : \\

\noindent The solution in this case is given by
\begin{eqnarray}
f(u, \rho ) &=& \frac{ \nu_{1} E_{0}^{2} (u) }{ m (m+c) } \left( \rho + \frac{ 1 }{2 (m+c)} \right) e^{- 2 m \rho } + \frac{ C_{1} (u)}{m (m+c)} e^{ - 2 m \rho } \nonumber \\ 
& & - \frac{2}{c} C_{2} (u) + C_{3} (u) e^{ 2 c \rho }  \, .
\label{pp_wave_3}
\end{eqnarray}     
Note that the condition $ \mu_{1} = - 2 m $ leads to the relation
\begin{eqnarray}
\mu^{2} ( \alpha \sigma + 1 )^{2} \left( \frac{ \mu ( \alpha \sigma + 1 )^{2}}{ 4 \alpha } - m \right)^{2} \left( \frac{ 1 }{4} \sigma^{2} + \frac{ \Lambda \alpha }{ \mu^{2} ( \alpha \sigma + 1 )^{2} } \right) \nonumber \\
= \left( m^{2} - \frac{ \mu^{2} ( \alpha \sigma + 1 )^{2} \sigma ( \alpha \sigma + 2 ) }{ 16 \alpha } \right)^{2}
\label{relation_2}
\end{eqnarray}
between the parameters. \\

\noindent {\bf 4.} The case $ \mu_{1} \neq 2 c $, $ \mu_{1} \neq - 2 m $, $ c = - m $ and $ m \neq 0 $ : \\

\noindent For this case, the solution can be obtained in the form
\begin{equation}
f ( u , \rho ) = - \frac{ 2 \bar{\nu}_{1} E_{0}^{2} (u) }{( \mu_{1} - 2 c ) c } \rho e^{ 2 c \rho } + \frac{ 4 C_{1} (u) }{ \mu_{1} ( \mu_{1} - 2 c )} e^{ \mu_{1} \rho } - \frac{ 2 }{c} C_{2} (u) + C_{3} (u) e^{ 2 c \rho }
\label{pp_wave_4}
\end{equation}
where $ \bar{\nu}_{1} $ is defined as
\begin{equation}
\bar{\nu}_{1} = \left. \nu_{1} \right \vert_{m=-c} = \frac{ \mu ( \alpha \sigma + 1 ) }{ 2 } \left( 1 - \frac{ 2 \alpha c }{ \mu ( \alpha \sigma + 1 ) } + \frac{ 2 \alpha^{2} }{ \mu^{2} ( \alpha \sigma + 1 )^{3} } ( c^{2} - \alpha \Lambda ) \right) \, .
\label{bar_nu}
\end{equation}
The condition $ c = - m $ requires the relation
\begin{equation}
\frac{ 1 }{4} \sigma^{2} + \frac{ \Lambda \alpha }{ \mu^{2} ( \alpha \sigma + 1 )^{2} } = \left( \frac{ \alpha m^{2} }{ \mu^{2} ( \alpha \sigma + 1 )^{3} } - \frac{ 1 }{2} \left( \sigma + \frac{ 2 \alpha^{2} \Lambda }{ \mu^{2} ( \alpha \sigma + 1 )^{3} } \right) \right)^{2}
\label{relation_3}
\end{equation}
between the parameters of the theory. \\

\noindent {\bf 5.} The case $ \mu_{1} \neq 2 c $ and $ m = 0 $ ; \\

\noindent This is the case where the Maxwell field is independent of the radial coordinate. Then the solution for this case can be given by
\begin{equation}
f(u, \rho) = \frac{ \nu_{0} E_{0}^{2} (u) }{ \mu_{1} c } \left( 2 \rho + \frac{ 1}{c} \right) + \frac{ 4 C_{1} (u) }{ \mu_{1} ( \mu_{1} - 2 c) } e^{ \mu_{1} \rho } - \frac{ 2}{c} C_{2} (u) + C_{3} (u) e^{ 2 c \rho } 
\label{pp_wave_5}
\end{equation}
where $ \nu_{0} $ is identified as 
\begin{equation}
\nu_{0} = \left. \nu_{1} \right \vert_{m=0} =  \frac{ \mu ( \alpha \sigma + 1 ) }{ 2 } \left( 1 + \frac{ 2 \alpha c }{ \mu ( \alpha \sigma + 1 ) } + \frac{ 2 \alpha^{2} }{ \mu^{2} ( \alpha \sigma + 1 )^{3} } ( c^{2} - \alpha \Lambda ) \right)  \, .
\label{nu_zero}
\end{equation}
\\
\noindent {\bf 6.} The case $ \mu_{1} = 2 c $ and $ m = 0 $ : \\

\noindent This is the another case where the Maxwell field is $ u $-dependent only. This special case leads to the solution 
\begin{equation}
f(u, \rho ) = \frac{ \nu_{0} E_{0}^{2} (u) }{ c^{2} } \left( \rho + \frac{ 1 }{ 2 c} \right) + \frac{ 2 C_{1} (u) }{ c } \rho e^{ 2 c \rho } - \frac{2}{c} C_{2} (u) + C_{3} (u) e^{ 2 c \rho } 
\label{pp_wave_6}
\end{equation}
provided that the parameters obey the relation (\ref{relation_1}). 

\section{ Noether charge in MMG with Maxwell-Chern-Simons coupling } 
\label{section_3}

In this section, by employing Noether-Wald formalism, we construct conserved charges in Minimal Massive Gravity coupled with Maxwell-Chern-Simons theory. To introduce formalism briefly, let us consider an $ n $-dimensional theory described by a Lagrangian $n$-form $ L $ such that the variation $ \delta L $ is induced by a $ p $-form field variation $ \delta \phi $ which can be written as 
\begin{equation}
\delta L = \delta \phi \wedge E( \phi ) + d \theta ( \phi , \delta \phi )
\label{variation_lagrangian}
\end{equation}
where $(n-1)$-form boundary term $ \theta (\phi , \delta \phi )$ can be identified as symplectic potential. Here $ E ( \phi ) $ denotes the equations of motion obtained from arbitrary variation. By considering anti-symmetrized variation of $ \theta $, one can also define the symplectic current $ \Omega $ in the form 
\begin{equation}
\Omega( \phi , \delta_{1} \phi , \delta_{2} \phi ) = \delta_{1} \theta ( \phi , \delta_{2} \phi ) - \delta_{2} \theta ( \phi , \delta_{1} \phi ) \, . \label{symplectic_current}
\end{equation}
Furthermore, one can define a symplectic form by integrating the symplectic current over a Cauchy surface $C$ as
\begin{equation}
\tilde{\Omega}  ( \phi , \delta_{1} \phi , \delta_{2} \phi ) = \int_{C} \Omega( \phi , \delta_{1} \phi , \delta_{2} \phi ) \, .
\label{symplectic_form}
\end{equation}
Let us suppose that the variation of the Lagrangian with respect to fields is induced by a diffeomorphism generated by a vector field $ \xi $ such that $ \delta_{ \xi } \phi = {\cal L}_{ \xi } \phi $. Note that $ {\cal L}_{ \xi } $ denotes the Lie derivative of $p$-form field $ \phi $. Then for the variation of the Lagrangian $n$-form, one can write
\begin{eqnarray}
\delta_{\xi} L &=& {\cal L}_{\xi} L  \nonumber \\ 
& = & d ( \iota_{\xi} L ) + \iota_{\xi} ( d L ) \nonumber \\
& = & d ( \iota_{\xi} L ) 
\end{eqnarray}
since $ d L $, being an $(n+1)$-form, vanishes for $n$-dimensional spacetime. Then if one defines $(n-1)$-form Noether current $ {\cal J}_{\xi} $ associated with vector field $ \xi $ in the form
\begin{equation}
{\cal J}_{\xi} = \theta ( \phi , \delta_{\xi} \phi ) - \iota_{\xi} L \, , 
\label{noether_current}
\end{equation}
one obtains
\begin{equation}
d {\cal J}_{\xi} = - \delta_{\xi} \phi \wedge E ( \phi ) \, .
\end{equation}
Obviously, $ d {\cal J}_{\xi} = 0 $ when $ E ( \phi ) = 0 $ (i.e on-shell). This further implies that one can define $ {\cal J}_{\xi} = d Q_{\xi} $ where $ Q_{\xi} $ can be identified as $ (n-2) $-form Noether charge. 
In the diffeomorphism method developed by Wald, on-shell variation $ \delta_{\xi} \phi $ of the fields corresponds to flow vector related to one-parameter family of diffeomorphisms generated by vector field $ \xi $. Then a Hamiltonian $ H_{\xi} $ that describes the dynamics generated by vector field $ \xi $ can be related to symplectic form by the relation
\begin{equation}
\delta H_{\xi} = \tilde{\Omega} ( \phi , \delta \phi , {\cal L}_{\xi} \phi )  = \int_{C} \Omega ( \phi , \delta \phi , {\cal L}_{\xi}  \phi ) \, . 
\label{variation_hamiltonian}
\end{equation} 
By setting $ \delta_{1} = \delta $, i.e arbitrary variation, and $ \delta_{2} = \delta_{\xi} $ in the expression (\ref{symplectic_current}) for the symplectic current, one obtains
\begin{eqnarray}
\Omega ( \phi , \delta \phi , \delta_{\xi} \phi ) &=& \delta \theta ( \phi , \delta_{\xi} \phi ) - \delta_{\xi} \theta ( \phi , \delta \phi ) \, .\nonumber \\
& = & \delta \theta ( \phi , \delta_{\xi} \phi ) - \iota_{\xi} d \theta - d ( \iota_{\xi} \theta ) \, . 
\end{eqnarray} 
where we have used
\begin{equation}
\delta_{\xi} \theta = {\cal L}_{\xi} \theta = \iota_{\xi} d \theta + d ( \iota_{\xi} \theta ) \, .
\end{equation}  
Next, we consider arbitrary variation of Noether current $ {\cal J}_{\xi} $. Assume that the vector field $ \xi $ is fixed such that its variation becomes $ \delta \xi = 0 $. Then
\begin{equation}
\delta {\cal J}_{\xi} = \delta \theta - \iota_{\xi} \delta L  \, .
\label{variation_noether_current}
\end{equation} 
Furthermore, if $ \delta \phi $ satisfies the linearized field equations, one can replace $ \delta {\cal J}_{\xi } $ by $ \delta ( d Q_{\xi} ) = d ( \delta Q_{\xi} ) $. Also using the variation of Lagrangian (\ref{variation_lagrangian}), one obtains 
\begin{equation}
\delta \theta = \delta ( d Q_{\xi} ) + \iota_{\xi} d \theta 
\end{equation}
where we consider that $ E (\phi) = 0 $ (i.e on-shell). Therefore,
\begin{eqnarray}   
\Omega ( \phi , \delta \phi , \delta_{\xi} \phi ) & = & \Omega ( \phi , \delta \phi , {\cal L}_{\xi} \phi ) \nonumber \\
& = & \delta \theta ( \phi , \delta_{\xi} \phi ) - \delta_{\xi} \theta ( \phi , \delta \phi ) \nonumber \\
& = & \delta ( d Q_{\xi} ) - d ( \iota_{\xi} \theta ) \nonumber \\
& = & d ( \delta Q_{\xi} - \iota_{\xi} \theta )  \, . 
\end{eqnarray}
Then $ \delta H_{\xi} $ becomes
\begin{eqnarray}
\delta H_{\xi} &=& \int_{C} d ( \delta Q_{\xi} - \iota_{\xi} \theta  ) \noindent \\
&=& \int_{ \partial C} ( \delta Q_{\xi} - \iota_{\xi} \theta ) \, .
\end{eqnarray}
where $ \partial C $ is the boundary of Cauchy surface. Then one can define Hamiltonian charge in the form \cite{iyer}
\begin{equation}
\delta H_{\xi} = \int_{ \partial C_{\infty}} ( \delta Q_{\xi} - \iota_{\xi} \theta ) 
\end{equation}
at spatial infinity $ \partial C_{\infty} $. For instance, let us consider a stationary black hole with the Killing field $ \xi $ that vanishes on a bifurcation surface $ {\cal B} $ (i.e over the horizon). Note that the symplectic current $ \Omega ( \phi , \delta \phi , {\cal L}_{\xi}  \phi ) $ vanishes if $ \xi $ is a symmetry of all dynamical fields in the theory. Then one obtains
\begin{equation}
\delta_{\xi} H = \oint_{ \partial C} ( \delta Q_{\xi} - \iota_{\xi} \theta ) = 0 \, , 
\label{boundary}
\end{equation}  
where in this case the boundary is considered to be closed. Also, if one considers that the Cauchy hypersurface $C$ possesses inner and outer boundaries being at bifurcation surface $ {\cal B} $ and spatial infinity $ \partial C_{\infty} $ respectively, then using (\ref{boundary}), one can write
\begin{equation}
\int_{\cal B} ( \delta Q_{\xi} - \iota_{\xi} \theta ) = \int_{\partial C_{\infty}} ( \delta Q_{\xi} - \iota_{\xi} \theta ) \, . 
\end{equation}
In addition, notice that since $ \xi $ vanishes over horizon, we have $ \iota_{\xi} \theta = 0 $ over $ {\cal B} $. As a result, one obtains
\begin{equation}
\int_{\cal B} \delta Q_{\xi} = \int_{\partial C_{\infty} } ( \delta Q_{\xi} - \iota_{\xi} \theta ) \, . 
\end{equation}
For a stationary black hole solution with time translational and axial rotational symmetries, we can take $\xi $ as the Killing vector expressed in the form $ \xi = \frac{ \partial }{\partial t } + \varOmega_{h} \frac{ \partial }{ \partial \varphi } $ where $ \varOmega_{h} $ corresponds to angular velocity of the horizon. In that case, up to a multiplicative constant, the energy and angular momentum can be defined as 
\begin{equation}
\delta {\cal E} = \int_{ \partial C_{\infty} } \, \delta Q^{\prime} \left[ \frac{ \partial }{ \partial t } \right] 
\label{mass}
\end{equation}
and
\begin{equation}
\delta J = - \int_{ \partial C_{\infty} } \, \delta Q^{\prime} \left[ \frac{ \partial }{ \partial \varphi } \right]
\label{angular_momentum}
\end{equation}
respectively where we identify
\begin{equation}
\delta Q^{\prime}_{\xi} = \delta Q_{\xi} - \iota_{\xi} \theta ( \phi , \delta \phi )
\label{charge_variation_form}
\end{equation}
as the charge variation forms. 

Now let us consider $3$-dimensional MMG theory coupled with Maxwell-Chern-Simons. Assuming that the variation of the total Lagrangian $3$-form $ L = L_{MMG} + L_{E} $ with respect to fields $ e^{a} $, $ \omega_{ab} $, $ \lambda_{a} $ and $ A $ is induced by a diffeomorphism generated by a vector field $ \xi $, we obtain
\begin{equation} 
\delta_{\xi} L = \delta_{\xi} e^{c} \wedge E_{c} + \delta_{\xi} \omega_{ab} \wedge E^{ab} + \delta_{\xi} \lambda_{a} \wedge N^{a} - \delta_{\xi} A \wedge N + d \theta ( \phi , \delta_{\xi} \phi )
\end{equation}
where $ E_{c} $, $ E^{ab} $, $ N^{a} $ and $ N $ denote the field equations (\ref{eqn_1_5}), (\ref{eqn_1_6}), (\ref{eqn_1_7}) and (\ref{maxwell_equation}) respectively and $ \phi $ denotes the collection of fields $ ( e^{a} , \omega^{ab} , \lambda^{a} , A ) $. The boundary term $ \theta ( \phi , \delta_{\xi} \phi ) $ reads
\begin{eqnarray}
\theta ( \phi , \delta_{\xi} \phi ) = - \frac{ \sigma }{2} \delta_{\xi} \omega_{ab} \wedge \ast ( e^{a} \wedge e^{b} ) + \frac{ 1 }{2 \mu} \delta_{\xi} \omega_{ba} \wedge \omega^{ab} \nonumber \\
+ \delta_{\xi} e^{a} \wedge \lambda_{a} - \delta_{\xi}  A \wedge \ast F - \frac{ 1 }{2} m \delta_{\xi}  A \wedge A \, . 
\label{diffeomorphism_boundary}
\end{eqnarray}  
In order to obtain conserved charge, we also need field variations 
\begin{eqnarray}
\delta_{\xi} e^{a} &=& {\cal L}_{\xi} e^{a} =  d ( \iota_{\xi} e^{a} ) + \iota_{\xi} d e^{a} \nonumber \\
& = & D ( \iota_{\xi} e^{a} ) - \iota_{\xi} \omega^{a}\,_{b} \wedge e^{b} + \iota_{\xi} T^{a}  \, ,
\end{eqnarray}
\begin{eqnarray}
\delta_{\xi} \omega^{ab} & = & {\cal L}_{\xi} \omega^{ab} =  d ( \iota_{\xi} \omega^{ab} ) + \iota_{\xi} d \omega^{ab} \nonumber \\
& = & \iota_{\xi} R^{ab} + D ( \iota_{\xi} \omega^{ab} ) 
\end{eqnarray}
where $ D $ denotes covariant exterior derivative with respect to connection $ \omega $. For the fields $ \lambda_{a} $ and $ A $, diffeomorphism induced variations can be expressed in the forms
\begin{equation}
\delta_{\xi} \lambda_{a} = {\cal L}_{\xi} \lambda_{a} =  d ( \iota_{\xi} \lambda_{a} ) + \iota_{\xi} d \lambda_{a} 
\end{equation}
and
\begin{equation}
\delta_{\xi} A  =  {\cal L}_{\xi} A  =  d ( \iota_{\xi} A ) + \iota_{\xi} d A
\end{equation}
respectively. Next we calculate $ {\cal J}_{\xi} = \theta ( \phi , \delta_{\xi} \phi ) - \iota_{\xi} L $ and obtain
\begin{equation}   
\theta ( \phi , \delta_{\xi} \phi ) - \iota_{\xi} L = d Q_{\xi} - \iota_{\xi} e^{c} \wedge E_{c} - \iota_{\xi} \omega_{ab} \wedge E^{ab} - \iota_{\xi} \lambda_{a} \wedge N^{a} + \iota_{\xi} A \wedge N
\end{equation}  
which implies that $ d {\cal J}_{\xi} = 0 $ on-shell. Here the Noether charge $ Q_{\xi} $ can be calculated as
\begin{eqnarray}
Q_{\xi} ( \omega ) & = & - \frac{ 1 }{ 2 \mu } \iota_{\xi} \omega_{ab} \wedge \omega^{ab} + \iota_{\xi} e^{c} \wedge \lambda_{c} - \iota_{\xi} A \wedge \left( \ast F + \frac { 1 }{2} m A \right) \nonumber \\
 & & - \frac{ \sigma }{2} \iota_{\xi} \omega_{ab} \wedge \ast ( e^{a} \wedge e^{b} ) \, .
 \label{noether_charge_1}
 \end{eqnarray}
Note that this is the general expression for the diffeomorphism-invariant Noether charge in MMG coupled with Maxwell-Chern-Simons theory. Note that a similar charge expression has been obtained also in \cite{nam_1} for MMG coupled with Maxwell-Chern-Simons theory where a similar formalism is used. We further remark that when $ \alpha = 0 $ and in the absence of Maxwell-Chern-Simons coupling, this expression reduces to charge expression in TMG presented in \cite{miskovic}. Then, in terms of torsion-free connection $ \bar{\omega}_{ab} $, $ Q_{\xi} $ can be obtained in the form
\begin{eqnarray}
Q_{\xi} ( \bar{\omega} ) & = & - \frac{ 1 }{2 \mu} \iota_{\xi} \bar{\omega}_{ab} \wedge \bar{\omega}^{ab} - \frac{ \alpha }{ 2 \mu} \epsilon_{abc} \left( \iota_{\xi} \bar{\omega}^{ab} \wedge \lambda^{c} + \iota_{\xi} \lambda^{c} \wedge \bar{\omega}^{ab} \right) \nonumber \\
& & + \frac{ \alpha^{2} }{\mu} \iota_{\xi} \lambda^{c} \wedge \lambda_{c} + \iota_{\xi} e^{c} \wedge \lambda_{c} - \iota_{\xi} A \wedge \left( \ast F + \frac{ 1 }{2} m A \right) \nonumber \\
& & - \frac{ \sigma }{2} \iota_{\xi} \bar{\omega}_{ab} \wedge \ast ( e^{a} \wedge e^{b} ) + \alpha \sigma \iota_{\xi} \lambda^{c} \wedge e_{c}
\label{noether_charge_2}
\end{eqnarray}      
where $ \lambda^{a} $ is given in (\ref{lambda_2}) in terms of Schouten one-form and Maxwell field. Then the charge variation forms $ \delta Q^{\prime}_{\xi} $ can be calculated in the form
\begin{eqnarray}
\delta Q^{\prime}_{\xi} ( \omega ) & = & \delta Q_{\xi} ( \omega ) - \iota_{\xi} \theta ( \phi , \delta \phi ) \nonumber \\
& = & - \frac{ 1 }{ \mu } \iota_{\xi} \omega_{ab} \wedge \delta \omega^{ab} + \iota_{\xi} e^{c} \wedge \delta \lambda_{c} - \iota_{\xi} A \wedge \left( \delta ( \ast F ) + \frac{ 1 }{2} m \delta A \right) \nonumber \\
&  & - \frac{ \sigma }{ 2 } \epsilon_{abc} \iota_{\xi} \omega^{ab} \wedge \delta e^{c} + \delta e^{c} \wedge \iota_{\xi} \lambda_{c} - \delta A \wedge \left( \iota_{\xi} ( \ast F ) + \frac{ 1 }{ 2 } m \, \iota_{\xi} A \right) \nonumber \\
&  & - \frac{ \sigma }{ 2 } \epsilon_{abc} \delta \omega^{ab} \wedge \iota_{\xi} e^{c} \, .
\label{charge_variation_form_1}
\end{eqnarray}  
In terms of torsion-free connection, the charge variation forms can be obtained as 
\begin{eqnarray}
\delta Q^{\prime}_{\xi} ( \bar{\omega} ) & = & - \frac{ 1 }{ \mu } \iota_{\xi} \bar{\omega}_{ab} \wedge \delta \bar{\omega}^{ab} - \frac{ \alpha }{\mu} \epsilon_{abc} \iota_{\xi} \bar{\omega}^{ab} \wedge \delta \lambda^{c} - \frac{ \alpha }{\mu} \epsilon_{abc} \iota_{\xi} \lambda^{c} \wedge \delta \bar{\omega}^{ab} \nonumber \\
& & + \frac{ 2 \alpha^{2} }{ \mu } \iota_{\xi} \lambda_{c} \wedge \delta \lambda^{c} + ( \alpha \sigma + 1 ) \iota_{\xi} e^{c} \wedge \delta \lambda_{c} \nonumber \\ 
& & + ( \alpha \sigma + 1 ) \iota_{\xi} \lambda_{c} \wedge \delta e^{c} - \iota_{\xi} A \wedge \left( \delta ( \ast F ) + \frac{ 1 }{ 2 } m \delta A \right) \nonumber \\
& & - \frac{ \sigma }{ 2 } \epsilon_{abc} \iota_{\xi} \bar{\omega}^{ab} \wedge \delta e^{c} - \delta A \wedge \left( \iota_{\xi} ( \ast F ) + \frac{ 1 }{ 2 } m \, \iota_{\xi} A \right) \nonumber \\
& & - \frac{ \sigma }{ 2 } \epsilon_{abc} \, \delta \bar{\omega}^{ab} \wedge \iota_{\xi} e^{c} \, .
\label{charge_variation_form_2}
\end{eqnarray} 
We notice that to calculate the conserved charges associated with vector field $ \xi $, the variation forms should be integrated at spatial infinity. Now, to obtain conserved charge for $pp$-wave solution, we closely follow the approach in \cite{nam_1} and \cite{nam_2} where in these works the conserved charges are obtained by calculating the charge variation forms for warped AdS solution. For the $pp$-wave metric, note that $ \xi = \frac{ \partial }{ \partial v } : = \xi_{v} $ is the Killing vector that is obviously null. To obtain the charge associated with the Killing vector, one also needs to calculate the variation forms for the co-frame and connections. Then, for the metric given in (\ref{metric}), the variation 1-forms for the co-frame and the connection can be calculated as
\begin{equation}
\delta e^{1} = 0 \, , \qquad \delta e^{0} = - \frac{ 1 }{ 2 \sqrt{2} } \delta f \, d u \, , \qquad \delta e^{2} = \frac{ 1 }{ 2 \sqrt{2} } \delta f \, d u
\end{equation}
and
\begin{eqnarray}
\delta \bar{\omega}^{02} = 0 \, \quad  \delta \bar{\omega}^{01} = - \left( \sqrt{2} \delta \beta + \frac{c}{ 2 \sqrt{2} } \delta f \right) du \nonumber \\ 
\delta \bar{\omega}^{12} = - \left( \sqrt{2} \delta \beta + \frac{c}{ 2 \sqrt{2} } \delta f \right) du  
\end{eqnarray} 
respectively. Note that variation of metric function $f(u, \rho ) $ (and variation of $ \beta ( u , \rho ) $) should be performed by considering the variation of wave profile functions $ C_{i} $ that appear in the solution since these parameters  behave as physical quantities. Then, it can be seen that the variation of co-frame and connection can be interpreted as difference between the solution and the background as is also indicated in \cite{nam_2}. On the other hand, the Maxwell field variations that we need can be presented as
\begin{equation}
\delta ( \ast F ) = - \sqrt{2} \delta E \, d u \, , \qquad \delta A = \frac{ \sqrt{2} }{m} \delta E_{0} \, e^{ - m \rho } \, d u
\end{equation}
for $ m \neq 0 $ where we have also used anti-self duality of the components of the Maxwell field. Note that variation of auxiliary field $ \lambda^{a} $ can be obtained similarly by using the variations for the co-frame, connection and Maxwell fields. Then if one uses variation forms for the fields, one can calculate the charge by using the expression (\ref{charge_variation_form_2}). After performing necessary calculations, one finds $ \delta Q^{\prime}_{v} = 0 $ for the charge variation associated with null Killing field $ \xi_{v} $ and as a result it becomes $ Q^{\prime}_{v} = 0 $ i.e a vanishing charge for $pp$-wave solution. This vanishing charge can be interpreted as the vanishing energy since the coordinate $ v $ can be identified as a time coordinate if one writes the metric in the form 
\begin{equation}
g = - \frac{ e^{4c \rho } }{ f } d v^{2} + f \left( d u + \frac{ e^{ 2 c \rho } }{ f } d v \right)^{2} + d \rho^{2} 
\end{equation}      
that is identical to ADM form. Indeed, comparing this form of $pp$-wave metric with the black hole solution (3.4) presented in \cite{clement}, one can see that the coordinate $ v $ is identical with time coordinate in the work \cite{clement}. We further remark that the result of vanishing charge of $pp$-wave solution completely agrees with the vanishing charge (in fact vanishing mass) of the black hole solution (3.4) with a null Killing vector in 3-dimensional massive gravity.

Before closing this section, it is also of great interest to construct Lorentz-diffeomorphism Noether charge in MMG coupled with Maxwell-Chern-Simons theory in this case by taking Lorentz gauge transformations together with diffeomorphisms into consideration. To obtain conserved charges for this case, one can consider that the field variations of Lagrangian are induced by a Lorentz-diffeomorphism generated by a vector field $ \xi $. This method has been recently introduced in \cite{jacobson} to calculate the entropy of a black hole as a Lorentz-diffeomorphism Noether charge. In that work, in order to prevent the divergence of connection field over the black hole horizon, the authors have proposed the idea of introducing Lorentz-Lie derivative in calculating the black hole entropy consistently. Later, the method has been extended to construct conserved charges in 3D Chern-Simons-like gravity theories in the absence of matter coupling \cite{adami_1,adami_2} and then recently it has been employed to obtain  
the conserved charges for a BTZ solution of MMG coupled to a (constant) scalar field \cite{adami_3,adami_4}. Mainly, the authors have constructed quasi-local conserved charges in MMG with scalar field coupling by using a method previously developed in \cite{kulkarni}. To construct Lorentz-diffeomorphism invariant Noether charge in MMG with Maxwell-Chern-Simons coupling, let us consider the Lorenz-diffeomorphism induced field variations of total Lagrangian generated by the vector $ \xi $. Then the variation of the Lagrangian can be written as 
\begin{equation} 
\bar{\delta}_{\xi} L = \bar{\delta}_{\xi} e^{c} \wedge E_{c} + \bar{\delta}_{\xi} \omega_{ab} \wedge E^{ab} + \bar{\delta}_{\xi} \lambda_{a} \wedge N^{a} - \bar{\delta}_{\xi} A \wedge N + d \theta ( \phi , \bar{\delta}_{\xi} \phi ) 
\label{lagrangian_lorentz_variation}
\end{equation}
where again $ E_{c} $, $ E^{ab} $, $ N^{a} $ and $ N $ denote the field equations (\ref{eqn_1_5}), (\ref{eqn_1_6}), (\ref{eqn_1_7}) and (\ref{maxwell_equation}) respectively and $ \phi $ denotes the collection of fields $ ( e^{a} , \omega^{ab} , \lambda^{a} , A ) $ as before. The boundary term $ \theta ( \phi , \bar{\delta}_{\xi} \phi ) $ reads
\begin{eqnarray}
\theta ( \phi , \bar{\delta}_{\xi} \phi ) = - \frac{ \sigma }{2} \bar{\delta}_{\xi} \omega_{ab} \wedge \ast ( e^{a} \wedge e^{b} ) + \frac{ 1 }{2 \mu} \bar{\delta}_{\xi} \omega_{ba} \wedge \omega^{ab} \nonumber \\
+ \bar{\delta}_{\xi} e^{a} \wedge \lambda_{a} - \bar{\delta}_{\xi}  A \wedge \ast F - \frac{ 1 }{2} m \bar{\delta}_{\xi}  A \wedge A \, . 
\label{lorentz_diffeomorphism_boundary}
\end{eqnarray}  
Note that all the barred terms involve Lorentz-diffeomorphism induced variations. In this case, the field variations become
\begin{eqnarray}
\bar{\delta}_{\xi} e^{a} &=& \bar{\cal L}_{\xi} e^{a} = {\cal L}_{\xi} e^{a} + \chi^{a}_{\xi \, b} \wedge e^{b} \nonumber \\  
& = & d ( \iota_{\xi} e^{a} ) + \iota_{\xi} d e^{a} + \chi^{a}_{\xi \, b} \wedge e^{b} \nonumber \\
& = & D ( \iota_{\xi} e^{a} ) + ( \chi^{a}_{\xi \, b} - \iota_{\xi} \omega^{a}\,_{b} ) \wedge e^{b} + \iota_{\xi} T^{a}  \, ,
\label{coframe_field_variation}
\end{eqnarray}
\begin{eqnarray}
\bar{\delta}_{\xi} \omega^{ab} & = & \bar{\cal L}_{\xi} \omega^{ab} - d \chi^{ab}_{\xi} \nonumber \\
& = & {\cal L}_{\xi} \omega^{ab} - D \chi^{ab}_{\xi} \nonumber \\
& = & d ( \iota_{\xi} \omega^{ab} ) + \iota_{\xi} d \omega^{ab} - D \chi^{ab}_{\xi} \nonumber \\
& = & \iota_{\xi} R^{ab} + D ( \iota_{\xi} \omega^{ab} ) - D \chi^{ab}_{\xi}
\label{connection_field_variation}
\end{eqnarray}
where $ \chi^{a}_{\xi \, b} $ represents the generators of Lorentz group $ SO(2,1) $. Note that $ \chi_{\xi}^{ab} = - \chi_{\xi}^{ba} $. Furthermore, generators $ \chi^{a}_{\xi \, b} $ do not depend on the fields but depend only on spacetime coordinates and vector field $ \xi $. In addition, $ \bar{\cal L}_{\xi} $ denotes Lorentz-Lie derivative operator while $ {\cal L}_{\xi} $ specifies the ordinary Lie derivative and $ D $ denotes covariant exterior derivative with respect to connection $ \omega $. On the other hand, the auxiliary field $ \lambda_{a} $ and Maxwell 1-form field $ A $ variations can be expressed in the forms
\begin{eqnarray}
\bar{\delta}_{\xi} \lambda^{a} & = & \bar{\cal L}_{\xi} \lambda^{a} = {\cal L}_{\xi} \lambda^{a} + \chi^{a}_{\xi \, b} \wedge \lambda^{b} \nonumber \\
& = &  d ( \iota_{\xi} \lambda^{a} ) + \iota_{\xi} d \lambda^{a} + \chi^{a}_{\xi \, b} \wedge \lambda^{b} \nonumber \\
& = & D ( \iota_{\xi} \lambda^{a} ) + \iota_{\xi} D \lambda^{a} + ( \chi^{a}_{\xi \, b} - \iota_{\xi} \omega^{a}\,_{b} ) \wedge \lambda^{b}
\label{lambda_field_variation}
\end{eqnarray}
and
\begin{equation}
\bar{\delta}_{\xi} A  =  \bar{\cal L}_{\xi} A  = {\cal L}_{\xi} A =  d ( \iota_{\xi} A ) + \iota_{\xi} d A 
\label{maxwell_field_variation}
\end{equation}  
respectively. Note that due to existence of gravitational Chern-Simons term in MMG Lagrangian, the theory becomes Lorentz non-covariant under Lorentz gauge tranformations (also see \cite{tachikawa}). Therefore Lorentz-diffeomorphism induced variation of Lagrangian $ L $ can be written as 
\begin{equation}
\bar{\delta}_{\xi} L = \bar{\cal L}_{\xi} L + d \Phi_{\xi} \, . 
\label{non_covariant_term_1}
\end{equation} 
In fact, one can obtain $ \Phi_{\xi} $ by calculating the Lorentz-Lie derivative of total Lagrangian and comparing with left hand side. Note that Lorentz-Lie derivative and exterior derivative operators do not commute except for Maxwell field. Then using commutator identities
\begin{equation}    
( \bar{\cal L}_{\xi} d - d \bar{\cal L}_{\xi} ) e^{a} = - d \chi^{a}_{\xi \, b } \wedge e^{b} \, ,
\label{commutator_coframe}
\end{equation}   
\begin{equation}
( \bar{\cal L}_{\xi} d - d \bar{\cal L}_{\xi} ) \omega^{ a b } = - D ( d \chi^{a b}_{\xi } )  \, ,
\label{commutator_connection}
\end{equation}    
\begin{equation}
( \bar{\cal L}_{\xi} d - d \bar{\cal L}_{\xi} ) \lambda^{a} = - d \chi^{a}_{\xi \, b} \wedge \lambda^{b}
\label{commutator_lambda}
\end{equation}
and
\begin{equation}
( \bar{\cal L}_{\xi} d - d \bar{\cal L}_{\xi} ) A = 0 
\label{commutator_maxwell_field}
\end{equation}
one can obtain $ \Phi_{\xi} $ in the form
\begin{equation}
\Phi_{\xi} = - \frac{ 1 }{ 2 \mu } d \chi^{ab}_{\xi} \wedge \omega_{ab} \, . 
\label{phi}
\end{equation}
Now, let us define the Noether current $ \bar{\cal J}_{\xi} $ in the form
\begin{equation}
\bar{\cal J}_{\xi} = \theta ( \phi , \bar{\delta}_{\xi} \phi ) - \iota_{\xi} L - \Phi_{\xi} \, .
\label{lorentz_noether_current}
\end{equation}
Then the straighforward calculation yields
\begin{equation}
\bar{\cal J}_{\xi} = d \bar{Q}_{\xi} - \iota_{\xi} e^{c} \wedge E_{c} - ( \iota_{\xi} \omega_{ab} - \chi^{\xi}_{ab} ) \wedge E^{ab} - \iota_{\xi} \lambda^{a} \wedge N_{a} + \iota_{\xi} A \wedge N
\end{equation}
which again implies that $ d \bar{\cal J}_{\xi} = 0 $ on shell. Notice that $ \bar{Q}_{\xi} $ denotes Lorentz-diffeomorphism Noether charge that can be written explicitly as
\begin{equation}
\bar{Q}_{\xi} ( \omega ) = Q_{\xi} ( \omega ) + \frac{\sigma}{2} \chi^{\xi}_{ab} \wedge \ast ( e^{a} \wedge e^{b} ) + \frac{ 1 }{ \mu } \chi^{\xi}_{ab} \wedge \omega^{ab} 
\label{lorentz_noether_charge_1}
\end{equation}
where $ Q_{\xi} (\omega ) $ is given in (\ref{noether_charge_1}). Note that this is the general expression for Lorentz-diffeomorphism invariant Noether charge in MMG coupled with Maxwell-Chern-Simons theory. In terms of torsion-free connection, it becomes 
\begin{equation}
\bar{Q}_{\xi} ( \bar{\omega} ) = Q_{\xi} ( \bar{\omega} ) + \frac{\sigma}{2} \chi^{\xi}_{ab} \wedge \ast ( e^{a} \wedge e^{b} ) + \frac{ 1 }{ \mu } \chi^{\xi}_{ab} \wedge \bar{\omega}^{ab} + \frac{ \alpha }{ \mu} \epsilon_{abc} \chi_{\xi}^{ab} 
\wedge \lambda^{c}  
\label{lorentz_noether_charge_2}
\end{equation}
where in this case $ Q_{\xi} ( \bar{\omega} ) $ is the Noether charge expression (\ref{noether_charge_2}) obtained in terms of torsion-free connection. Next, let us set $ \delta_{1} = \delta $ and $ \delta_{2} = \bar{\delta}_{\xi} $ in the symplectic current expression (\ref{symplectic_current}) and define
\begin{equation}
\bar{\Omega} ( \phi , \delta \phi , \bar{\delta}_{\xi} \phi  ) = \delta \theta ( \phi , \bar{\delta}_{\xi} \phi ) - \bar{\delta}_{\xi} \theta ( \phi , \delta \phi )
\label{lorentz_symplectic_current} 
\end{equation}
as the new symplectic current 2-form that involves Lorentz-diffeomorphism induced variations. Again we notice that due to the existence of gravitational Chern-Simons term in the Lagrangian which makes it non-covariant under Lorentz-diffeomorphisms, one can write
\begin{equation}
\bar{\delta}_{\xi} \theta ( \phi , \delta \phi ) = \bar{\cal L}_{\xi} \theta ( \phi , \delta \phi ) + \Xi_{\xi} \, . 
\label{non_covariant_term_2}
\end{equation}   
Then by evaluating Lorentz-Lie derivative of the boundary term with arbitrary variations and comparing with the left hand side, one can obtain the term $ \Xi_{\xi} $ in the form
\begin{equation}
\Xi_{\xi} = - \frac{ 1 }{ 2 \mu } d \chi^{\xi}_{ab} \wedge \delta \omega^{ab} \, .
\label{xi}
\end{equation} 
Next, we consider arbitrary variations of Noether current $ \bar{\cal J}_{\xi} $. Notice that $ \delta \bar{\cal J}_{\xi} = \delta ( d \bar{Q}_{\xi} ) = d ( \delta \bar{Q}_{\xi} ) $ on shell and $ \delta \xi = 0 $. Then using
\begin{eqnarray}
\bar{\cal L}_{\xi} \theta ( \phi \delta \phi ) & = & {\cal L}_{\xi} \theta ( \phi , \delta \phi ) \nonumber \\
& = & d ( \iota_{\xi} \theta ( \phi , \delta \phi ) ) + \iota_{\xi} ( d \theta ( \phi , \delta \phi ) )
\end{eqnarray}
and expression (\ref{non_covariant_term_2}), one finally obtains symplectic current in the form
\begin{equation}
\bar{\Omega} ( \phi , \delta \phi , \bar{\delta}_{\xi} \phi  ) = d ( \delta \bar{Q}_{\xi} - \iota_{\xi} \theta ( \phi , \delta \phi ) ) + \delta \Phi_{\xi} - \Xi_{\xi} \, . 
\end{equation} 
Note that in general $ \delta \Phi_{\xi} - \Xi_{\xi} = d \Pi_{\xi} $ \cite{tachikawa}. Now using the expressions (\ref{phi}) and (\ref{xi}) and also noting that $ \delta \chi_{\xi} = 0 $, one obtains 
\begin{equation}
\delta \Phi_{\xi} - \Xi_{\xi} = d \Pi_{\xi} = 0 \, .
\end{equation}
Then choosing $ \Pi_{\xi} = 0 $, the symplectic current finally becomes
\begin{equation}
\bar{\Omega} ( \phi , \delta \phi , \bar{\delta}_{\xi} \phi  ) = d ( \delta \bar{Q}_{\xi} - \iota_{\xi} \theta ( \phi , \delta \phi ) ) \, .
\end{equation} 
Note that in that case one can define the new charge variation forms as
\begin{equation}
\delta \bar{Q}^{\prime}_{\xi} =  \delta \bar{Q}_{\xi} - \iota_{\xi} \theta ( \phi , \delta \phi ) \, .
\end{equation} 
Finally a straightforward calculation yields
\begin{equation}
\delta \bar{Q}^{\prime}_{\xi} ( \omega ) = \delta Q^{\prime}_{\xi} (\omega)  + \frac{ 1 }{ \mu } \chi^{\xi}_{ab} \wedge \delta \omega^{ab} + \frac{\sigma}{2} \epsilon_{abc} \chi_{\xi}^{ab} \wedge \delta e^{c}
\label{lorentz_charge_variation_form_1}
\end{equation}
in terms of connection $ \omega $ and
\begin{equation}
\delta \bar{Q}^{\prime}_{\xi} ( \bar{\omega} ) = \delta Q^{\prime}_{\xi} ( \bar{\omega} ) + \frac{ 1 }{ \mu } \chi^{\xi}_{ab} \wedge \delta \bar{\omega}^{ab} + \frac{ \alpha }{ \mu } \epsilon_{abc} \chi_{\xi}^{ab} \wedge \delta \lambda^{c} + \frac{\sigma}{2} \epsilon_{abc} \chi_{\xi}^{ab} \wedge \delta e^{c}
\label{lorentz_charge_variation_form_2}
\end{equation}
in terms of torsion-free connection where $ \delta Q^{\prime}_{\xi} (\omega) $ and $ \delta Q^{\prime}_{\xi} ( \bar{\omega} ) $ are given in (\ref{charge_variation_form_1}) and (\ref{charge_variation_form_2}) respectively. Obviously the integration of charge variation forms at spatial infinity yield conserved charges associated with the vector field $ \xi $. Also remark that when $ \xi $ is a Killing vector generators $ \chi_{\xi}^{ab} $ can be obtained by demanding that Lorentz-Lie derivative of co-frame vanishes, i.e $ \bar{\cal L}_{\xi} e^{a} = 0 $ \cite{jacobson}. Then when $ \xi $ is a Killing vector, one obtains
\begin{equation}
\bar{\cal L}_{\xi} e^{a} = \iota_{\xi} T^{a} - \iota_{\xi} \omega^{a}\,_{b} \wedge e^{b} + D ( \iota_{\xi} e^{a} ) + \chi^{a}_{\xi \, b} \wedge e^{b} = 0 \, .
\end{equation}
Now using
\begin{equation}
T^{a} = K^{a}\,_{b} \wedge e^{b}
\end{equation}
and decomposition of connections in the form
\begin{equation}
\omega^{a}\,_{b} = \bar{\omega}^{a}\,_{b} + K^{a}\,_{b} \, ,
\end{equation}
one finds that
\begin{equation}
\chi^{a}_{\xi \, b } = \iota_{\xi} \bar{\omega}^{a}\,_{b} + \frac{ 1 }{ 2 } \left( \iota^{a} \bar{D} ( \iota_{\xi} e_{b} ) - \iota_{b} \bar{D} ( \iota_{\xi} e^{a} ) \right)  
\label{lorentz_generator}
\end{equation}
where $ \bar{D} $ denotes covariant exterior derivative with respect to torsion-free connection $ \bar{\omega} $. Then if one calculates $ \chi^{a}_{\xi \, b } $ for the $ pp $-wave solution with Killing vector $ \xi = \xi_{v} = \frac{ \partial}{ \partial v } $, one finds that $ \chi^{a}_{ \xi b } = 0 $ i.e all the components vanish. Therefore, we conclude that Lorentz-diffeomorphism Noether charge also vanishes for $ pp $-wave solution.

\section{Conclusion} 

In this work, we have obtained a family of $pp$-wave solutions of MMG minimally coupled to Maxwell-Chern-Simons theory. To this end, by using the language of differential forms we have first derived the governing field equations by making variations of the matter-coupled MMG Lagrangian with respect to co-frame and connection 1-forms, Maxwell potential 1-form as well as the auxiliary 1-form that determines the torsion. Then by solving auxiliary 1-form from the field equations, we have obtained the MMG gravitational field equation with the effective source 2-form that contains quadratic terms in the stress-energy 2-forms associated with the Maxwell field as well as their covariant derivative. As was mentioned and explicitly shown in \cite{arvanitakis_2} and \cite{cebeci}, the source term of MMG completely differs from the source term of TMG where for the latter the source is just expressed by the stress-energy 2-form $ \tau_{c} $ associated with the matter fields. Furthermore, unlike TMG, the covariant exterior derivative of the source term does not vanish but it has to satisfy a consistency relation given in (\ref{consistency}).  
Then, by proposing a $ pp $-wave ansatz for the metric and a convenient ansatz with two components for Maxwell field, we have explicitly derived the field equations. By making an elaborate analysis of the field equations, we have explicitly shown that MMG minimally coupled with Maxwell-Chern-Simons admits $pp$-wave solutions provided that the electric and the magnetic components of Maxwell 2-form field obey an anti-self-duality condition. In fact, by using both MMG and Maxwell field equations, we have explicitly shown that an anti-self duality relation should exist between the electric and magnetic components of Maxwell 2-form field. Next, by employing Noether-Wald formalism, we have constructed diffeomorphism-invariant Noether charges in MMG with Maxwell-Chern-Simons coupling. We have also generalized the construction by obtaining Lorentz-diffeomorphism Noether charges where in this we take Lorentz gauge transformations together with diffeomorphisms into consideration. By using the general expression for conserved charges, we have found a vanishing charge for $ pp $-wave solution with null Killing vector.  

As a future research, we hope that within exterior algebra formalism one can examine stationary black hole solutions of MMG coupled with Maxwell-Chern-Simons theory and calculate Lorentz-diffeomorphism Noether charges. It would also be of great interest to investigate the matter-coupled solutions in some other 3d massive gravity models presented in the works \cite{afshar_1,ozkan,afshar_2,geiller}.  

\section*{Acknowledgments }  

\noindent We would like to thank \"{O}zg\"{u}r Sar{\i}o\u{g}lu for useful comments and discussions at the early stages of this work.

\appendix

\section{Connection, Curvature, Schouten and Cotton forms for the $pp$-wave metric}
For the $pp$-wave metric, 
\begin{equation}
g = f (u, \rho) d u^{2} + 2 e^{ 2 c \rho } du \, dv + d \rho^{2}
\label{pp} 
\end{equation}
we may choose co-frame 1-forms in the form
\begin{eqnarray}
e^{0} & = & \frac{ 1 }{\sqrt{2}} \left( \left( 1 - \frac{f}{2} \right) d u - e^{2c \rho } d v \right) \, , \quad e^{1} = d \rho \nonumber \\ 
e^{2} & = & \frac{ 1 }{\sqrt{2}} \left( \left( 1 + \frac{f}{2} \right) d u + e^{2c \rho } d v \right)  \, .
\label{coframe}
\end{eqnarray}
Then using torsion-free Cartan structure equations
\begin{equation}
d e^{a} + \bar{\omega}^{a}\,_{b} \wedge e^{b} = 0
\end{equation} 
one can determine the Levi-Civita connection 1-forms as 
\begin{equation}
\bar{\omega}^{0}\,_{1} = (c-\beta) e^{0} - \beta e^{2} \, , \quad \bar{\omega}^{0}\,_{2} = c e^{1} \, , \quad \bar{\omega}^{1}\,_{2} = - \beta e^{0} - ( \beta + c) e^{2}  \, .
\label{connection}
\end{equation}  
in terms of the function $ \beta=\beta(u,\rho)$ defined as 
\begin{equation}
\beta (u, \rho) = \frac{ 1 }{4} f^{\prime} - \frac{c}{2} f
\end{equation}
where a prime denotes differentiation with respect to metric variable $ \rho $. Next, using second Cartan structure equations 
\begin{equation}
\bar{R}^{ab} = d \bar{\omega}^{ab} + \bar{\omega}^{a}\,_{c} \wedge \bar{\omega}^{cb}
\end{equation}
the curvature 2-forms associated with the Levi-Civita connection can be obtained in the form 
\begin{eqnarray}
\bar{R}^{01} & = & - \beta^{\prime} e^{1} \wedge e^{2} + (c^{2} - \beta^{\prime}) e^{1} \wedge e^{0} \nonumber \\
\bar{R}^{02} & = & - c^{2} e^{0} \wedge e^{2} \, , \quad \bar{R}^{12} = ( - \beta^{\prime} - c^{2} ) e^{1} \wedge e^{2} - \beta^{\prime} e^{1} \wedge e^{0}  \, . 
\label{curvature}
\end{eqnarray}
Next, the corresponding torsion-free Schouten 1-forms can be calculated from
\begin{equation}
\bar{Y}^{a} = \bar{P}^{a} - \frac{ 1 }{4} \bar{R} e^{a}
\end{equation}
where $ \bar{P}^{a} = \iota_{b} \bar{R}^{ba} $ and $ \bar{R} = \iota_{a} \bar{P}^{a} $ denote torsion-free Ricci-1-forms and scalar curvature respectively. Then Schouten 1-forms read as 
\begin{equation}
\bar{Y}^{1} = - \frac{1}{2} c^{2} e^{1} \, , \quad 
\bar{Y}^{0} = ( \beta^{\prime} - \frac{ 1 }{ 2 } c^{2} ) e^{0} + \beta^{\prime} e^{2}  \, , \quad
\bar{Y}^{2} = - \beta^{\prime} e^{0} + ( - \beta^{\prime} - \frac{1}{2} c^{2} ) e^{2}  \, . 
\end{equation} 
Finally, from $ \bar{C}^{a} = \bar{D} \bar{Y}^{a} $, Cotton 2-forms can be computed in the form
\begin{equation}
\bar{C}^{1} = 0 \, , \quad \bar{C}^{0} = ( \beta^{ \prime \prime} - c \beta^{\prime} ) e^{1} \wedge e^{2} + ( \beta^{ \prime \prime} - c \beta^{\prime} ) e^{1} \wedge e^{0} \, , \quad \bar{C}^{2} = - \bar{C}^{0}  \, .
\end{equation}

\end{document}